# Multi-scale approach to invasion percolation of rock fracture networks

Ali N. Ebrahimi, Falk K. Wittel*, Nuno A.M. Araújo, Hans J. Herrmann

*Computational Physics for Engineering Materials, Institute for Building Materials, ETH Zurich, Stefano-Franscini-Platz 3, CH-8093 Zurich, Switzerland*
*corresponding author: Tel. +41 44 633 2871. E-mail address: fwittel@ethz.ch*

## Abstract

A multi-scale scheme for the invasion percolation of rock fracture networks with heterogeneous fracture aperture fields is proposed. Inside fractures, fluid transport is calculated on the finest scale and found to be localized in channels as a consequence of the aperture field. The channel network is characterized and reduced to a vectorized artificial channel network (ACN). Different realizations of ACNs are used to systematically calculate efficient apertures for fluid transport inside differently sized fractures as well as fracture intersection and entry properties. Typical situations in fracture networks are parameterized by fracture inclination, flow path length along the fracture and intersection lengths in the entrance and outlet zones of fractures. Using these scaling relations obtained from the finer scales, we simulate the invasion process of immiscible fluids into saturated discrete fracture networks, which were studied in previous works.

## 1. Introduction

Transport of fluids in fractured geological media plays an important role in different applications such as subsurface hydrology, hydrocarbon recovery from natural reservoirs, safe storage facilities for captured $CO_2$ or hazardous wastes (Sahimi, 1995; Adler et al, 2013). The prediction of the penetration of an immiscible fluid into a fully saturated fracture network and its percolation is among the most challenging topics in sub-surface hydrology. Fractured rock, instead of behaving like an equivalent continuum, is only sparsely fractured, restricting fluid flow to a small part of the connected fractures (Reeves, et al., 2008). Fluids can take multiple pathways, can be trapped, and exhibit path instabilities and scale dependencies, what makes the prediction of effective properties rather challenging. Detailed descriptions however are rarely capable of capturing more than a handful of interconnected fractures and fail in representing the complex system dynamics emerging from the interaction of a huge number of locally interacting connected fractures. A partial solution to this dilemma was found in the Discrete Fracture Network (DFN) approach that assumes fluid flow to be entirely localized in the network of connected fractures (Smith and Schwartz, 1984; Huseby et al., 2001). DFN models have been widely applied to study fluid flow and transport characteristics of several fractured rocks with low permeability and fracture density (Reeves, et al., 2008; Sisavath, et al., 2004; Renshaw, 1999). As one expects, transport properties of the networks are strongly affected by the fracture density, size and interactions (Khamforoush and Shams, 2007; Koudina, et al., 1998; Dreuzy, et al., 2000). For a detailed description of the physics of fracture networks we refer to the reviews by Berkowitz (2002), Sahimi (1995), and Adler et

al., (2013).

Since fluid flow in a single fracture is at the bottom of any DFN model, a detailed description of its geometry and hydraulic behavior is essential (Charmet, et al., 1990; Pompe, et al., 1990; Adler et al., 2013). Realistic topological measurements of natural fracture planes revealed spatial correlations (Candela et al., 2012). From those, aperture field distributions can be obtained (Kumar, et al., 1997) and applied for studying the hydraulic behavior of the single fracture (Konzuk and Kueper, 2004). The most striking observation is that significant parts of the natural fractures have zero aperture since they are contact zones, while the fluid flow is naturally localized in areas with non-zero aperture which are also called open zones. Hence, the fluctuations in the aperture field lead to fluid flow in a network of channels (Tsakiroglou, 2002; Katsumi, et al., 2009) and fracture interactions are governed by the characteristics of these two networks. To model the displacement of immiscible fluids inside DFN models, invasion percolation (IP) theory proved to be an efficient tool (Wilkinson and Willemsen, 1983). In fact, IP is a modified form of ordinary percolation (Broadbent and Hammersley, 1957), only with a well-defined sequence of invasion events. In IP theory both gravity and viscous effects are neglected and only capillary forces are considered. However, IP is a valid approximation to describe the slow immiscible displacement of two phase flow in fracture networks.

In this work, details of the single fracture with its inherent disorder are upscaled to the fracture network by a two-step coarse graining method. Vectorized artificial channel networks (ACN) are used, that are highly compressed representations of channelized flow on the single fracture scale obtained from Finite Element (FE) simulations with heterogeneous aperture fields. ACNs are used to calculate the scaling behavior of hydraulic transport properties of the single fractures in terms of size dependent equivalent apertures, as well as size and angle dependent fracture-fracture interactions for entry apertures. The numerically obtained rules are then incorporated into a fracture-network consisting of 4000 intersecting fractures for a modified invasion percolation (MIP) simulation with two-phase flow (Wettstein et al., 2012). This way we obtain a more realistic physical description of the invasion process with coarse grained information from the fracture scale.

## 2. Methodology

In this methodological section, first flow at the fracture scale is addressed by FE simulations and as prerequisite for data compression for the ACN. As a next step we describe fracture interaction parameters, effective properties and their consideration on the DFN scale in trapping MIP.

### *2.1. Flow at the fracture scale*

Single fractures have fundamental meaning in fracture network modeling (Cappa, 2011) since from this, scale size effects emerge. The effect of fracture aperture and roughness on hydraulic properties was already addressed from the theoretical and numerical perspectives in the past (Drazer and Koplik, 2000; Drazer and Koplik, 2002; Auradou, et al., 2001; Berkowitz, 2002; Sahimi, 1995). In general, fracture surfaces exhibit spatial correlation (Méheust and Schmittbuhl, 2001) and are self-affine with roughness (Hurst) exponent close to 0.8 for diverse materials (Bouchaud, et al., 1990; Ansari-Rad, et al., 2012; Dyer, et al., 2012). However also roughness exponents close to 0.6 can be observed in the direction of slip at laboratory scale samples (Amitrano and Schmittbuhl, 2002) and even at the scale of natural faults (Bistacchi, et al., 2011; Candela, et al., 2012). In principle, fractures with a constant fracture opening have constant aperture and could be reduced to two parallel plates with Hagen-Poiseuille flow as well as the cubic law (Zimmerman and Bodvarsson, 1996; Konzuk and Kueper, 2004). However huge fluctuations of the aperture field arise due to contact zones between two facing fracture surfaces (Brown, 1987; Dreuzy, et al., 2012; Katsumi, et al., 2009). Detailed explanations of self-affine fracture surfaces and aperture field distributions can be found for example in Adler et al. (2013), Brown (1995) and Hamzehour et al., (2009). Isotropic self-affine fracture surfaces are constructed using a spectral method based on fractional Brownian motion (Schrenk, et al. (2013)). In a consecutive step a copy of the surface is vertically and horizontally displaced with respect to the original, resembling homogeneous shear displacement. The local aperture is then taken as distance between surfaces while overlapping and contacting regions are assigned zero apertures (Auradou, et al. (2005); Durham (1997); Katsumi, et al. (2009)).

In this study, uncorrelated random disordered fracture surfaces and hence uncorrelated random disordered aperture fields are used for generality, since there is no focus on a peculiar geological formation. However any type of aperture field can be treated analogously. The utilized disorder is given by a power-law probability density distribution such that the aperture $h_i$ of each site $i$ can be generated as $h_i=\exp(B(x_i-1))$ with the disorder parameter $B$ and $x_i$ being a uniformly distributed random number in [0,1] (Oliveira, et al., 2011; Braunstein et al., 2002). For each realization, an aperture $h_i$, $i=1,...,L^d$, (where $L$ is the size of the aperture field, e.g. of a square lattice and $d$ is the dimension of the system) is assigned to each site according to the mentioned power-law distribution. The number of cells is chosen with respect to the characteristic length $l$ for the aperture field that is assumed to be unit length. Between neighboring sites, values are interpolated by bi-cubic interpolation (see Fig.1(a)).

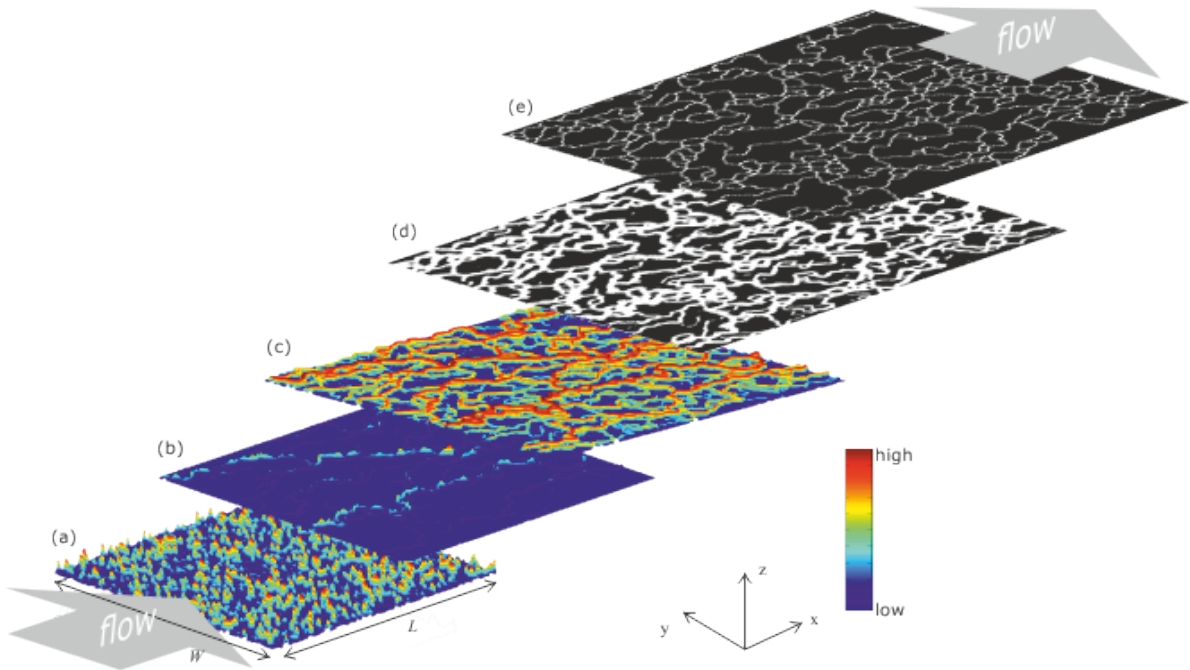

**Fig. 1. Coarse graining procedure: (a) Aperture field, (b) flow rate field, (c) equalized flow rate field, (d) binarized flow and (e) skeletonized field for a single disordered fracture.**

For sufficiently permeable fractures, single-phase flow is assumed for single fractures and hydraulic properties can be approximated by a parallel plate fracture (Méheust and Schmittbuhl, 2001). To model fluid flow inside of single fractures, the lubrication approximation, hence Stokes flow for small Reynolds number (Re<1), as well as negligible gravity (bond number<<1) and inertial forces are assumed. The Stokes equation between two parallel plates relates the pressure field $p$ to the velocity field $\upsilon$ by

$$\nabla p = \mu \Delta \upsilon \qquad (1)$$

with the dynamic viscosity $\mu$ of fluid. For a parallel plate the velocity $\upsilon$ only depends on plate distance ($z$-direction) while the pressure $p$ is only dependent on the $x$-direction. Hence Eq. 1 can be rewritten as

$$\frac{dp}{dx} = \mu \frac{d^2 \upsilon}{dz^2} = C, \qquad (2)$$

where $C$ is a constant (Méheust and Schmittbuhl, 2001). By solving Eq. 2, one gets the well-known "cubic law", which gives a good estimate of the volumetric flow rate $q$ through a single fracture as a function of the pressure gradient $\nabla p$ in the flow direction. A cubic law equation is given by

$$q = W \frac{h^3}{12 \mu L} \nabla p, \qquad (3)$$

where $h$ is the plate distance or fracture aperture and $W$ denotes the width of the fracture perpendicular to the flow direction (see Fig. 1(a)).

To simulate fluid flow through a rough fracture, an aperture field has to be used. Using the lubrication approximation, pressure $p$ only depends on the $x$- and $y$-direction, leading to the following form of the Stokes equation:

$$\nabla p(x, y) = \mu \frac{d^2 \upsilon(x, y, z)}{dz^2}. \qquad (4)$$

In order to simplify Eq. 4, the local direction $\hat{u}(x, y)$ and the local coordinate $u$ along this direction for the pressure gradient are introduced by

$$\frac{dp}{d\hat{u}} = \mu \frac{d^2 \upsilon(x, y, z)}{dz^2}. \qquad (5)$$

In order to reduce the problem to a two dimensional system, the local flow rate direction $\hat{q}(x, y)$ can be obtained by integration along the z direction

$$\hat{q}(x, y) = \int_{z_l}^{z_u} \upsilon(x, y, z) dz = \frac{\hat{h}(x, y)^3}{12 \mu} \nabla p, \qquad (6)$$

with $z_l$ and $z_u$ being the lower and upper surface heights respectively, hence the local cubic law with the aperture field $\hat{h}(x, y)$. For this two-dimensional approach, the main flow direction is assumed to be the $x$-direction with constant pressure at the inlet $p_{in}$ and outlet $p_{out}$. Also no-flow conditions are defined on boundaries parallel to the flow direction. Note that all quantities are kept in a dimensionless form by using the characteristic units of the numerical simulation, hence the characteristic length $l$ of the aperture field has length unity.

The FE method (FEM) is chosen to discretize Eq. 6 using the COMSOL Multiphysics® software, with the Matlab interface. The aperture field is discretized using structured quadrilateral elements of mesh size *d*. The weak form of the cubic law is obtained by the linear test functions and by applying Green's theorem (integration by parts) over the domain of the element. Since Gauss quadrature is used for the numerical integration, node values represent extrapolated values from the integration points. The system is solved by a MUltifrontal Massively Parallel Sparse Direct Solver (MUMPS). In order to apply the mentioned boundary conditions, aperture values at any boundary with no flow conditions are assumed to be zero and the apertures of inlet and outlet are treated like a big channel with the maximum aperture.

## 2.2. *Identification and generation of the artificial channel network (ACN)*

Since FEM simulations are computationally expensive, considering the required resolution and number of realizations for one value of disorder, a coarse graining strategy is worked out which is based on the strong channelization of fluid flow inside fractures and network analysis. First channel networks are extracted by using image analysis algorithms on the normalized flow rate fields. In a second step, several properties are extracted from the flow rate fields along the identified network skeletons. The channel identification algorithm and the successive construction of the artificial channel network is sketched in Fig.2 and described in the following.

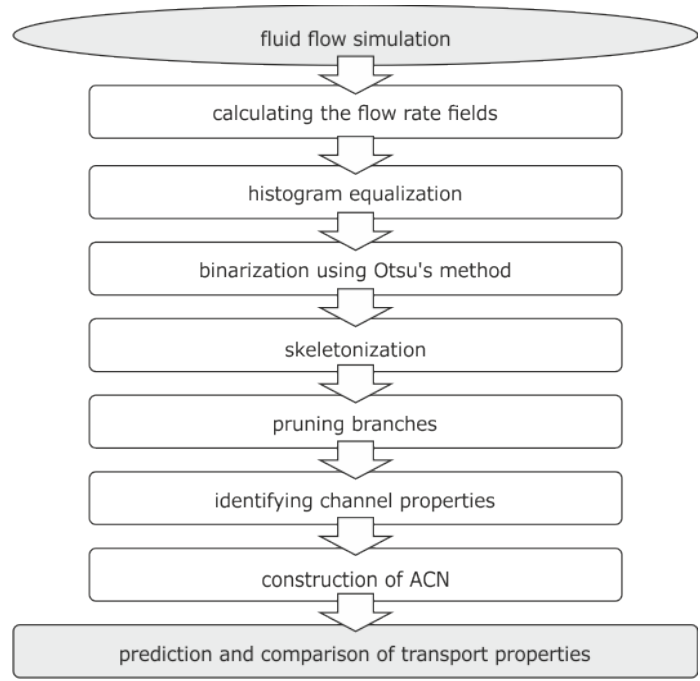

**Fig. 2. Algorithm to identify the channels and to create the artificial channel network.**

In order to identify the channels from the normalized flow rate field, the dynamic range of the intensity-level of the flow rate matrix is increased by a histogram equalization technique from the MATLAB Image Processing Toolbox (Gonzalez, et al., 2004). Fig. 1(c) demonstrates the significant improvement in the visibility of channels. To extract the skeleton of the network, channels are segmented by binarization using the Otsu's method (see Fig. 1(d)). Morphological operations on the binary images, like the medial axis transformation, reduce the network to a system of identical topology but of only one pixel with. In a next step unconnected branches and spurs with low impact on the total flow rate are pruned by repetitive elimination of the pixels with coordination number one. The pruned and skeletonized network is shown in Fig. 1(e). It is the starting point for extracting the topological properties of the network such as channels length, width and degree of connectivity. For this purpose the networks need to be separated into interconnected channels, what can be obtained by simply removing pixels with more than two connections from the skeleton. For further evaluation, removed pixels are stored as connection nodes. The degree of connectivity for each node is obtained by counting the number of the connections of each node to other channels. Note that channels shorter than the characteristic length of the aperture field are merged with neighboring larger channels. Since channels are now separated objects, they are labeled and further analyzed. Each channel gets evaluated by measuring its equivalent aperture, width and length, as well as the degree of their connectivity. These properties are measured along and in the vicinity of the channel directly from the field output of the FE simulation. Details on the procedure for calculating the equivalent apertures and widths of channels are given in Appendix A.

Based on extracted channel and network characteristics, an algorithm is developed to generate a reasonable, coarse grained ACN. This highly compressed representation of the fracture flow field will be used to numerically quantify the effect of system size on effective fracture apertures, effective entry apertures for fracture intersections and disorder averaged over many realizations. In principle many different construction principles for the ACN can be used alike. We base our algorithm on vectorizable random lattices (Mourkarzel and Herrmann, 1992). The construction of the ACN starts on a square lattice with $N_a$ nodes and characteristic size $a$ (rounded value of $\rho^{-0.5}$) that is in principle obtained from the extracted node density $\rho$ (Section 3.2). The tortuosity of channels is irrelevant, and only the distance between two nodes of channels, $d_{min}$ represented in Fig. 6 is used for the construction of ACN. Due to the merging condition mentioned above, the connected nodes in the ACN model cannot be closer than the minimum distance constraint between two connected channels $d_{min}$ obtained from the distribution of distance between two nodes. This condition is imposed by shrinking each grid cell to the size $b=a-d_{min}$ with fixed center. The positions of nodes are set inside these cells. In the next step, the degree of connectivity for each node is randomly distributed from the previously measured degree of node connectivity of the channels. Each node can only be connected to the nearest neighborhood nodes. Finally, the equivalent aperture and channel width is assigned to each connection, sampling from the previously numerically evaluated distribution functions.

*2.3.    Properties of fracture intersections*

Experimental results point at the critical role of the correct treatment of fracture intersections on the dynamics and transport pathways in invasion and transport processes (Ji, et al., 2006; Glass and LaViolette, 2004; Smith and Schwartz, 1984; Wood et al., 2005). Water is transported quickly along the fracture and then slows down at the intersections of fractures until the capillary pressure reaches the invasion threshold of the next fracture. Therefore, the intersections between fractures behave as capillary barriers that control the flow or invasion rate into the fracture network (Glass et al., 2003). Intersections of fractures can be regarded as ACNs interacting along intersection lines. Important properties are the equivalent aperture between two intersections, as well as the entry aperture e.g. from fracture i to fracture j. They depend on the characteristics of the underlying channel networks with disorder B, the intersection length Lij and distance dik, as well as their relative spatial orientation (see Fig. 3). In this work, we refrained from trying to find relations of general validity and focused on a numerical approach that calculates intersections of different ACNs instead.

In Fig. 3 flow through fracture *j* from fracture *i* to fracture *k* is sketched. The original equivalent aperture $h_j^{eq}$ of reference fracture *j* must be modified by considering the path length along the fracture $d_{ik}^j$, as well as the two lengths of the fracture intersection lines at the inlet $L_{ij}$ and outlet $L_{jk}$ as well as the contact angles between the fractures ($\theta$ and $\phi$). Considering the geometrical configuration of fracture *j*, $h_j^{eq}$ and $w_j$ are modified at the intersection, based on the rotation angles by the following equations:

$$\hat{h}_j^{eq} = h_j^{eq}/\sin(\theta); \qquad \hat{w}_j = w_j/\sin(\phi). \qquad (7)$$

The total flow rate from fracture *i* to fracture *k* is obtained by summing up the flow rates over all involved channels in the intersection line of fracture *j* and *k*.

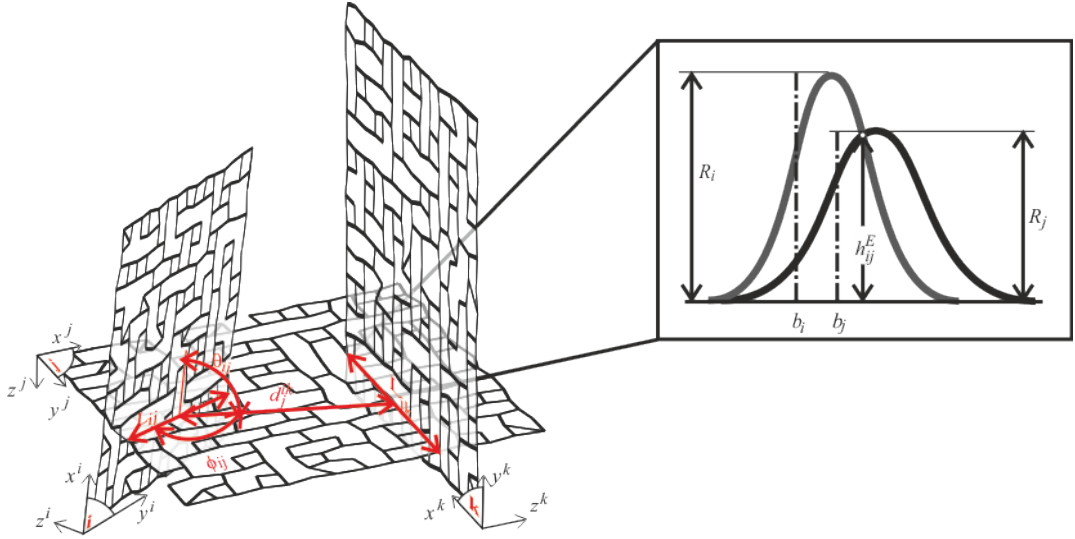

**Fig. 3. Schematic representation of the interaction between three fractures (*i*,*j*,*k*) and schematic cross section of the intersection between two channels (inset).**

In order to obtain effective entry apertures of fracture intersections, it is assumed that channels of one fracture only contribute to fluid transport into an intersecting fracture, if channels geometrically intersect. To study possible intersects between two channels we assume that channel *i* is in the reference plane, connected to a channel *j* that is rotated by the Euler angles $\theta$ and $\phi$ around the local *y* and *z* axes. The rotation around the *x* axis is considered by the length of the overlap between channels *i* and *j*. For critical rotation angles <10°, instantaneous invasion of geometrically overlapping channels is assumed. For angles >10°, the entry aperture of the single channel $h_{ij}^E$ can be obtained from the intersection point between two channels in the reference plane. As will be explained in Sec. 3.1, based on the detailed fracture simulations, the equivalent aperture size $h_j^{eq}$ and width $w_j$ of channels follow a Gaussian profile. Therefore for the intersecting channels, $h_{ij}^E$ in the intersection is calculated from the height of the intersect point (see inset in Fig. 3).

Once all entry apertures for overlapping channels are calculated from the total flow, the entry aperture for fracture *j* in the intersection of fracture *i*, is obtained. The effective entry aperture must be calculated for a wide range of possible fracture intersections and averaged over many realizations, before they are employed in the large fracture network simulation.

### 2.4. *Discrete Fracture Network Model*

The aim of this work is to improve the trapping MIP of DFNs previously developed by (Wettstein et al., 2012). The DFN construction is explained in detail by Huseby et al. (2001) and Khamforoush and Shams (2007). Briefly, the fracture network is composed of planar, regular or irregular, polygonal fractures. The fractures are oriented around the mean direction. For the sake of simplicity, each fracture is modeled by an individual polygon with $N_v$ vertices inscribed in a disk with radius $R_D$. The radii of the mentioned circles are randomly chosen in the interval [$r_{min}$, $r_{max}$], and the number of vertices $N_v$ is chosen in the interval [3, $N_{max}$], both from uniform distributions. The vertices are uniformly directed on the perimeter of the circle with uniformly distributed angles from [0,2$\pi$]. All fractures have centers that are uniformly distributed in the simulation box of size $L_N$. The orientation of each fracture is adjusted by the unit normal vector of its plane. In order to generate anisotropic fracture networks, the fracture normal vectors are distributed according to the Fisher distribution. In the standard cylindrical coordinates $\theta' \in [0,\pi]$ and $\varphi' \in [0,2\pi]$, the Fisher distribution of a random set of normal vectors is defined with a probability density function

$$f(\theta',k) = \frac{k}{4\pi \sinh k} \sin\theta' \exp(k\cos\theta'), \quad (8)$$

where k>0 is the dispersion parameter about the mean direction which controls the anisotropy in the DFN model. Note that Eq. 8 represents the particular case where initial coordinates $\theta'_0$ and $\varphi'_0$ are equal to zero. The distribution is rotationally symmetric around the initial mean direction, which coincides with the *z*-axis. Increasing the value of *k* concentrates the distribution around this axis. The distribution is unimodal for *k*>0 and is uniform on the sphere for *k*=0. In order to build a spanning cluster over the entire domain of the DFN, the fracture density of the system should be high enough. The detailed analysis for the effects of anisotropy and fracture density for the DFN model on the number of intersections of fractures is presented by Huseby et al. (2001) and Khamforoush et al. (2008). For the invaded DFN model, the global system matrix is composed of invaded fractures as elements and the connections as nodes using a simple channel model following Cacas et al. (1990) and Wettstein et al. (2012). After the DFN is constructed, the invasion of immiscible fluids can be simulated. Note that our algorithm should more correctly be named trapping MIP, since it considers entirely encircled non-invaded regions using a burning algorithm (Herrmann et al., 1984). After pruning the mentioned traps and dead ends from the percolating cluster, the invasion percolation backbone is used to calculate the flow through the remaining hydraulic network. The entry zone is chosen in the *xz*-plane on the left side at *y*=0, and the exit zone correspondingly as the *xz* -plane on the right side at *y*=$L_N$. Periodic boundaries are applied on the other four sides. A detailed description of the algorithm can be found in Wettstein et al. (2012).

## 3. Coarse graining calculations

Up-scaling in general goes along with loss or compression of information. Hence it is important to quantify the accuracy of the procedures. First a fixed degree of disorder is chosen for demonstration purposes. The distribution of channel network characteristics is studied in detail, before they are used to construct artificial ACNs. After verifying their behavior, parametric studies of intersecting networks are performed to obtain quantitative scaling relations for improving DFNs.

### *3.1. Quantification of channel properties*

The main assumption of the proposed approach is that flow inside fractures is localized within a channel network with distinct topological characteristics. The channels are created due to the dispersion of apertures and can been identified as flow pathways. In the past, numerical and experimental studies have explained the impact of these channels on the permeability of the single fracture at both experimental and field scales (Ishibashi, et al., 2009; Watanabe, et al., 2008; Talon, et al., 2010; Talon and Auradou, 2010). Therefore we solve the cubic law to obtain pressure fields that are used to calculate flow rate fields for specific realizations of two-dimensional aperture fields with disorder, defined by the power-law probability density distribution by parameter $B$ (see Sec.2.1.). The underlying two-dimensional aperture field has a characteristic length $l$ of unit length and a size of $L$ x $W$ of 80 x 80. Fluid of a dynamic viscosity $\mu$ of $10^{-4}$ enters at a pressure $p_{in} = 10^{-6}$ and leaves the system at zero pressure $p_{out} = 0$. We discretize the system by a square grid of element edge length $d$. To assess the mesh quality, a convergence study of decreasing $l/d$ was well-fitted to a power law with an exponent of 2.025 of convergence. Reasonable accuracy is obtained for meshes with values of about $d=0.25$ and smaller. The parameters are rendered dimensionless by the characteristic units of the numerical simulation. In order to use the cubic law, all dimensionless parameters are set to have a Reynolds number lower than one.

The emergence and impact of channelization strongly depends on the degree of disorder $B$ of the aperture field. Consequently the characteristic sizes and topologies of networks change as well. Hence, first we monitor the effect of the disorder on the ratio of the channelized flow rate compared to total flow in the entire fracture, as well as the open zone ratio (see Fig.4). By increasing heterogeneity of the aperture field $B$ from zero, one shifts from a parallel plate flow to strongly localized flow in channels that form a network with characteristic properties of their own, due to the increase of contact zones. In this way, the transition from weak to strong disorder system can be identified with respect to the system size (Andrade, Jr, et al., 2009).

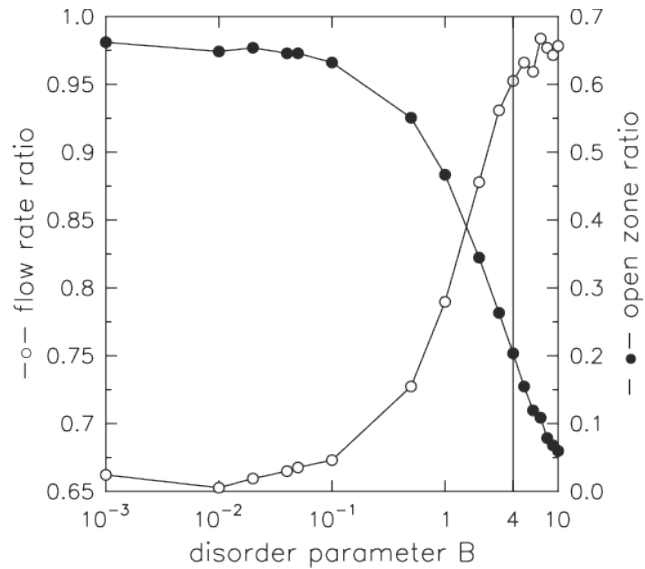

**Fig. 4. Effect of disorder on the fraction of channelized flow as well as the ratio of the open zone area.**

This study is limited to the case of $B=4$, when approximately 94% of the flow is channelized. In fact, this high impact of the channels on the total flow rate allows us to assume flow to be entirely localized within the network of channels. Note that a lower concentration of channels could be considered as a parallel plate flow parallel to the channel network. Channelization of the flow for the different realizations of aperture fields allows us to characterize the channel network topology to identify underlying length scales for scaling arguments of transport properties. The two main properties for channels are distribution functions for their width $w$, as well as their equivalent aperture $h^{eq}$. Using the described identification procedure for the network (Sec.2.2) we obtain information on all channels and hence normalized distribution functions $f$ (see Fig. 5):

$$f(h^{eq}) = 0.014 \exp\left(\frac{-(h^{eq}-0.0091)^2}{2\times 0.0301^2}\right), \quad f(w) = 0.014 \exp\left(\frac{-(w-1.3902)^2}{2\times 0.1988^2}\right). \quad (9)$$

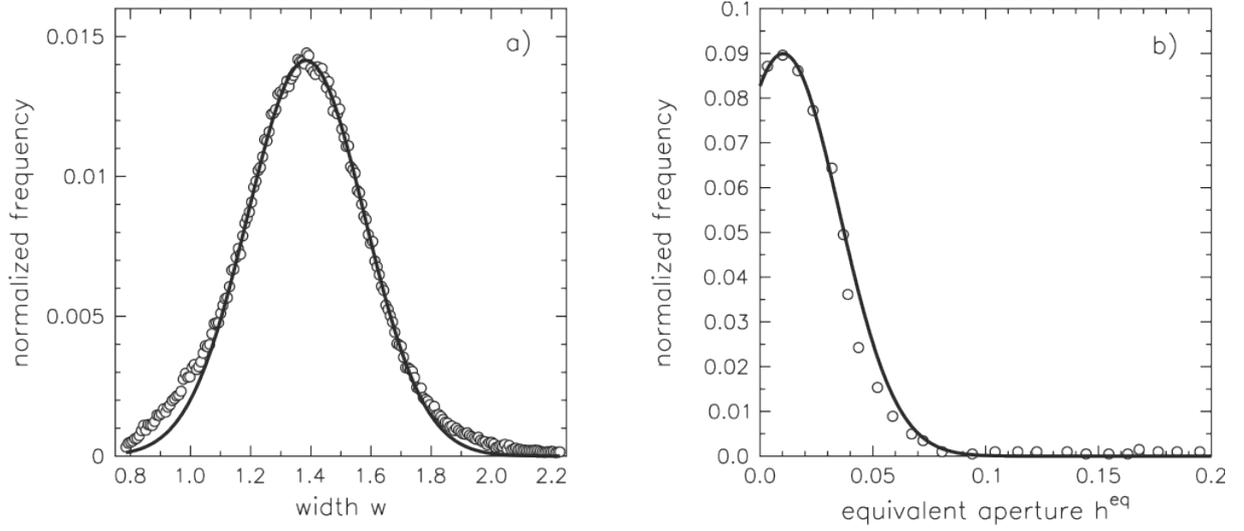

**Fig. 5.** The distribution of properties of channels, (a) width, and (b) equivalent size for a disordered system ($B$=4). Solid lines show the Gaussian fit (Eq. 9) of the measured values (circles).

### 3.2. *Characterization and validation of ACNs*

To be able to construct an ACN, topological features, such as node density, degree of connectivity and channel length need to be measured from the skeletonized networks. The node density is defined as a number of nodes $N_a$ per area of the fracture $N_a/L^2$. In this study, its value for the disordered system with $B$=4 is $\rho$=0.70±0.01 and the degree of connectivity for nodes are distributed between 3, 4 and 5 connections with 81%, 16% and 3%, respectively. The resulting distribution of channel lengths, hence the distance distribution between two connected nodes, can be compared to the original fracture, showing good agreement (see Fig. 6).

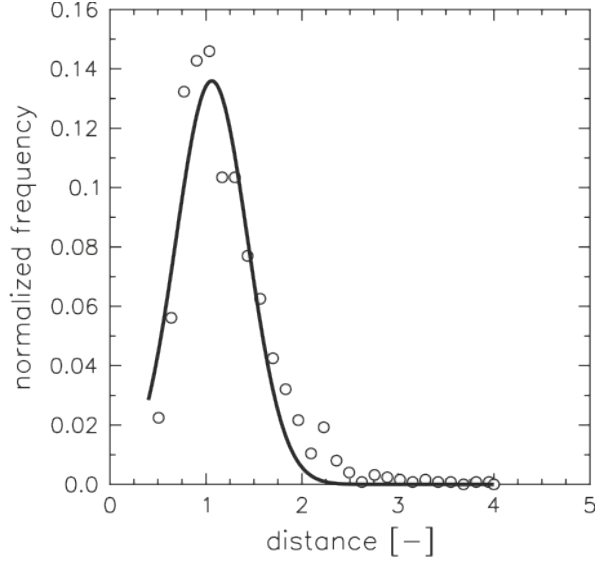

**Fig. 6. Distribution of the distances between interconnected nodes for the ACN (circles) and the original fracture (line).**

A much better validation is the comparison of pressure and flow rates between ACN and FE results. To do so, the cubic law is applied to each channel. Hence, mass balance for a typical node $i$ is represented by

$$\sum_{j=1}^{z} G_{i,j}\left(p_i - p_j\right) = 0 \text{ with } G_{ij} = w_{ij}\frac{(h_{ij}^{eq})^3}{12\mu l_{ij}}, \quad (10)$$

where $z$ denotes the degree of connectivity for node $i$, $w_{ij}$, $l_{ij}$, and $h_{ij}$ are width, length and equivalent aperture of the channel connecting node $i$ and $j$. Identical boundary conditions and fluid properties are taken as for the FE calculation. The set of linear equations is solved for all nodes to obtain the pressure at the nodes. The total flow inside the ACN is about 8% smaller than the one of the detailed fracture. This difference is largely explained by the fact that only flow in channels was considered that was quantified before (Fig. 4) to be ~94% of the total one for $B=4$. The equivalent fracture aperture is one of the physically important parameters that should be accurately predicted from the ACN model. Here the relative error between the equivalent aperture of the detailed fracture and ACN was calculated to be ~11%, what can be considered as small with respect to the huge data reduction by the ACN and hence performance gain.

### 3.3. *Parameterization of fracture interactions*

The two important effective properties to consider in the discrete fracture network are the equivalent aperture for flow between two fracture intersections and the entry aperture of intersections. Using the procedures described in Sec. 2.3., we obtain equivalent apertures with respect to the length of the intersections ($L_{ij}$, $L_{jk}$) path length $d_{ik}^{j}$ and rotation angle $\phi$ (see example in Fig. 7 for specific values).

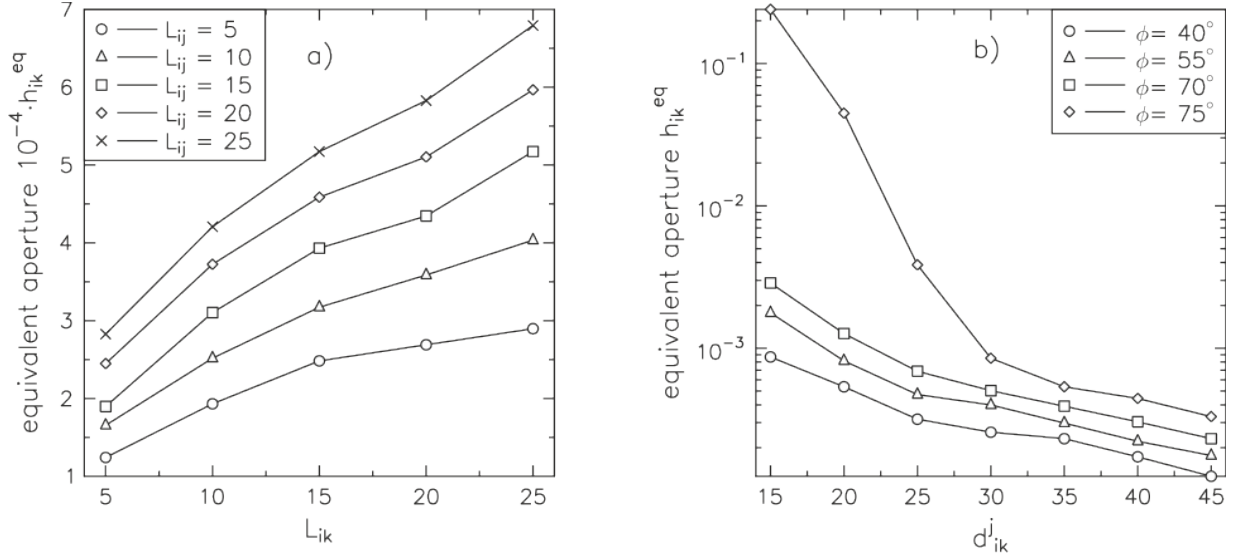

**Fig. 7. Equivalent aperture size over the length of the fracture intersections (a) with constant rotation angle $\phi$ and path length ($\phi\approx 60°$ and $d_{ik}\approx 35$) and (b) with constant intersection length ($L_{ij}=L_{jk}\approx 20$).**

To fit the simulation results, the quadratic dependency on the length of the intersections is considered by the expression

$$h_{ik}^{eq} = a_1^{eq} + a_2^{eq}(L_{ij}) + a_3^{eq}(L_{jk}) + a_4^{eq}(L_{ij})^2 + a_5^{eq}(L_{jk})^2. \qquad (11)$$

The parameters $\phi$ and $d_{ik}^{j}$ are involved in the fitting parameters $a_n^{eq}$ through

$$a_n^{eq} = b_{n1}^{eq} + b_{n2}^{eq}\sin(\varphi_{ij}) + b_{n3}^{eq} d_{ik}^{j}, \qquad (12)$$

where $b_{ni}^{eq}$, $i=1,2,3$ and $n=1,2,\ldots,5$ are the fitting parameters summarized in Tab. 1.

**Tab. 1. Fitting parameters for Eq. 12 ($B$=4).**

| $n$ | $b_{n1}^{eq}$ | $b_{n2}^{eq}$ | $b_{n3}^{eq}$ |
|---|---|---|---|
| 1 | -9.997e-04 | -4.225e-03 | 2.371e-03 |
| 2 | -1.884e-04 | 7.102e-04 | -1.123e-04 |
| 3 | 1.416e-04 | 7.73e-04 | -3.973e-04 |
| 4 | 3.344e-05 | -6.91e-05 | -1.48e-06 |
| 5 | -2.90e-06 | -1.66e-05 | 8.702e-06 |

For large angles $\phi$, fracture $i$ and $k$ can directly intersect due to different rotations, intersection and path lengths. This causes a sudden increase in the equivalent aperture value that is visible in Figure 7b for $\phi =75°$. However this case is more straightforwardly solved by considering direct connection between fracture $i$ and $k$ once the criterion $½L_{ij}\geq d_{ik}^{j}\cdot\cos(\phi)^{-1}$ for two connecting fractures is exceeded.

To obtain the value of the entry aperture $h_{ij}^{E}$ as function of the intersection length of the fractures $L_{ij}$, averages over 1000 realizations of the ACNs are made (see Fig. 8).

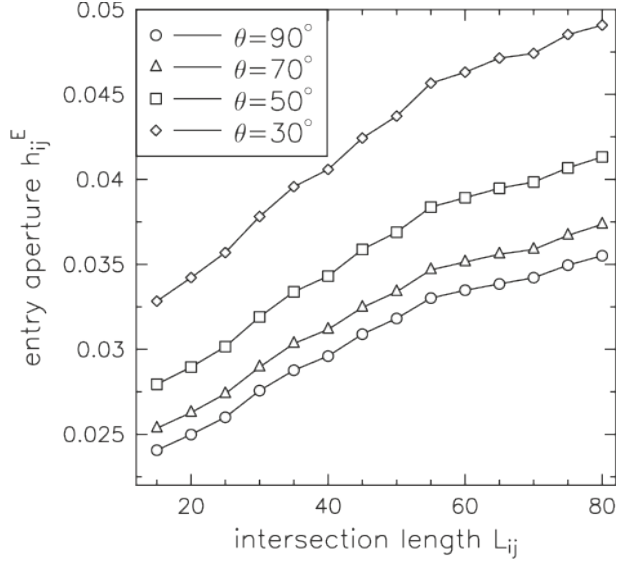

**Fig. 8. Dependence of the entry aperture on intersection length and angle $\theta$.**

The entry aperture can be fitted with reasonable accuracy by using a quadratic dependency on the length of the intersection, expressed by

$$h_{ij}^{E} = a_1^{E} + a_2^{E}(L_{ij}) + a_3^{E}(L_{ij})^2, \quad (13)$$

where $a_i^{E}$, $i=1,2,3$, are the fitting parameters for the entry aperture which are assumed to have a linear dependence on the rotation angle around $y$-axis with fitting parameters $b_i^{E}$, $i=1,2,3$ given in Tab. 2:

$$a_n^{E} = b_{n1}^{E} + b_{n2}^{E}\sin(\theta). \quad (14)$$

The dependency of the rotation angle around the $z$-direction is included in the interaction of the two channels and ignored in the calculation of the entry aperture due to its negligible impact.

**Tab.2. Fitting parameters for Eq. 14 ($B=4$).**

| $n$ | $b_{n1}^{E}$ | $b_{n2}^{E}$ |
|---|---|---|
| 1 | 2. 22e-2 | -3.65e-3 |
| 2 | 3.68e-4 | -5.70e-5 |
| 3 | -1.60e-6 | 1.71e-7 |

## 4. Invasion percolation of DFNs

At the scale of the single fracture and hence channel network, the complexity of the fluid transport is caused by the heterogeneous geometry and channelization as previously explained. At the network scale however, the complex behavior originates from the fracture size distribution and density, as well as size effects of fracture intersections that was quantified in Sec. 3.3. Only few researches have attempted to model the hydraulic behavior of the single fracture and the fracture network scale, simultaneously. In fact, this multi-scale analysis is essential due to the strong effects of both scales on the transport properties of fractured rocks (Hamzehpour, et al., 2009; Dreuzy, et al., 2012). In the following, we revisit the previously studied problems of two-phase invasion percolation (IP) in the form of immiscible displacement through isotropic and anisotropic DFNs (Wettstein et al., 2012). In contrast to the previous modifications, fracture intersections are now including the effect fracture orientation, intersection length, as well as the path length along the fractures by using the numerically found scaling relations for *B*=4.

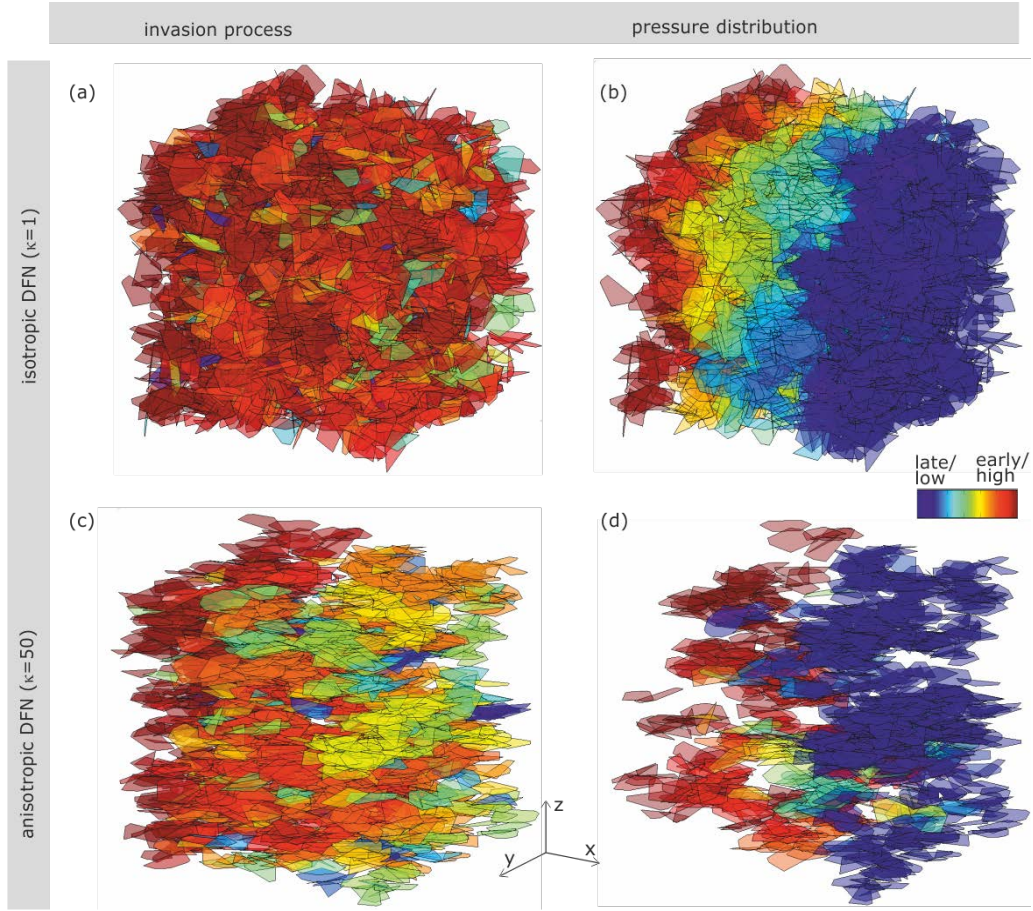

**Fig. 9. DFN model for full invasion with trapping MIP. (a) and (c) invaded fractures during invasion process. Colors indicate the invasion sequence. (b) and (d) show the pressure distribution in the flow fracture network.**

The network consists of $N_{fr}$=4000 fractures inside a cube of length $L_N$=10. For the DNF construction lower and upper bounds of radius of polygons are set to $r_{min}$=0.5, $r_{max}$=1 respectively, while the number of vertices is limited to $N_{max}$=8. These properties are similar with respect to the local fracture porosity and fracture area, only the fracture orientation is changed by setting the Fisher dispersion parameter $\kappa$=1 for the isotropic case (Figs. 9-11(a),(b)) and $\kappa$=50 for the highly anisotropic one (Figs. 9-11(c),(d)). The invasion process starts at the entry zone and progresses in the direction of the outlet (*x*-direction). After breakthrough, the invasion process is continued to mimic for example an infinite gas production rate at the inlet until no further fractures can be invaded. Note that in the network scale, flow inside the single fracture is assumed to be non-viscous (zero capillary number) and therefore as soon as the capillary pressure reaches the aperture threshold of the corresponding fracture, the entire fracture is instantaneously invaded. In order to obtain pressure distributions for the DFN model, a one phase flow calculation is performed analogous to the ACN model on the hydraulic network.

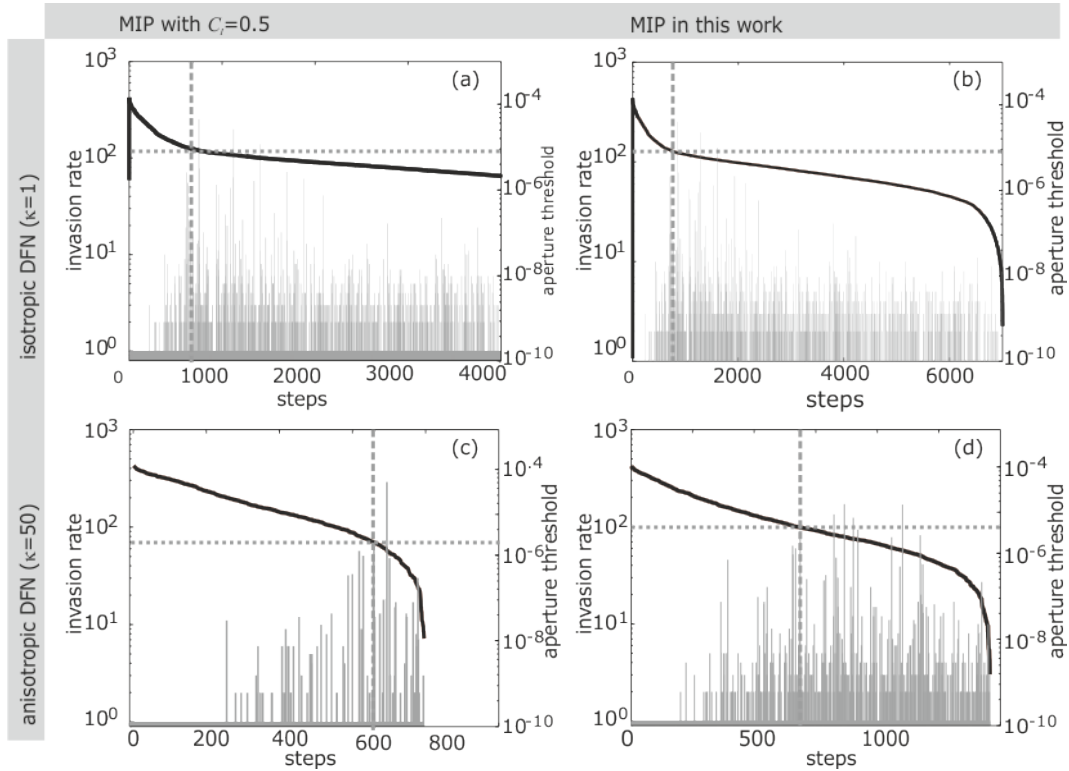

**Fig. 10. Invasion process with aperture threshold and invasion rate for each step for trapping MIP on the isotropic DFN ($\kappa=1$) and anisotropic DFN ($\kappa=50$). (a) and (c) trapping MIP ($C_I = 0.5$). (b) and (d) improved MIP. The breakthrough step is marked by a dashed line for the invasion steps and with a dotted line for the aperture threshold value.**

The drop of the aperture threshold for each invasion step characterizes the dynamics of the invasion process. The comparison between trapping MIP ($C_I = 0.5$; adjustment parameter for the inclination effect of intersections see (Wettstein et al. 2012)) and the newly developed MIP is presented in Figs. 10 and 11. The number of steps needed for a full invasion for the isotropic and anisotropic cases is significantly increased by the multi-scale procedure. This can be explained by a higher disorder for equivalent apertures, leading to smaller avalanches and hence more steps to invade all non-trapped fractures. The number of steps required to reach the percolation breakthrough, as well as the aperture threshold are more or less identical. Note that decreasing aperture threshold means increasing the of entry pressure.

The capillary pressure is calculated using the Young-Laplace equation and displayed in relation to the water saturation in Fig. 11, also referred to as water retention curve. As visible, the water saturations in the new MIP at the full invasion process for both isotropic and anisotropic cases are smaller than in the previous MIP model. This again is due to a higher degree of disorder, leading to the participation of channels with small entry and equivalent apertures as well as the lack of the connectivity of the anisotropic DFN, resulting in an increasing number of invasion steps for the multiscale approach.

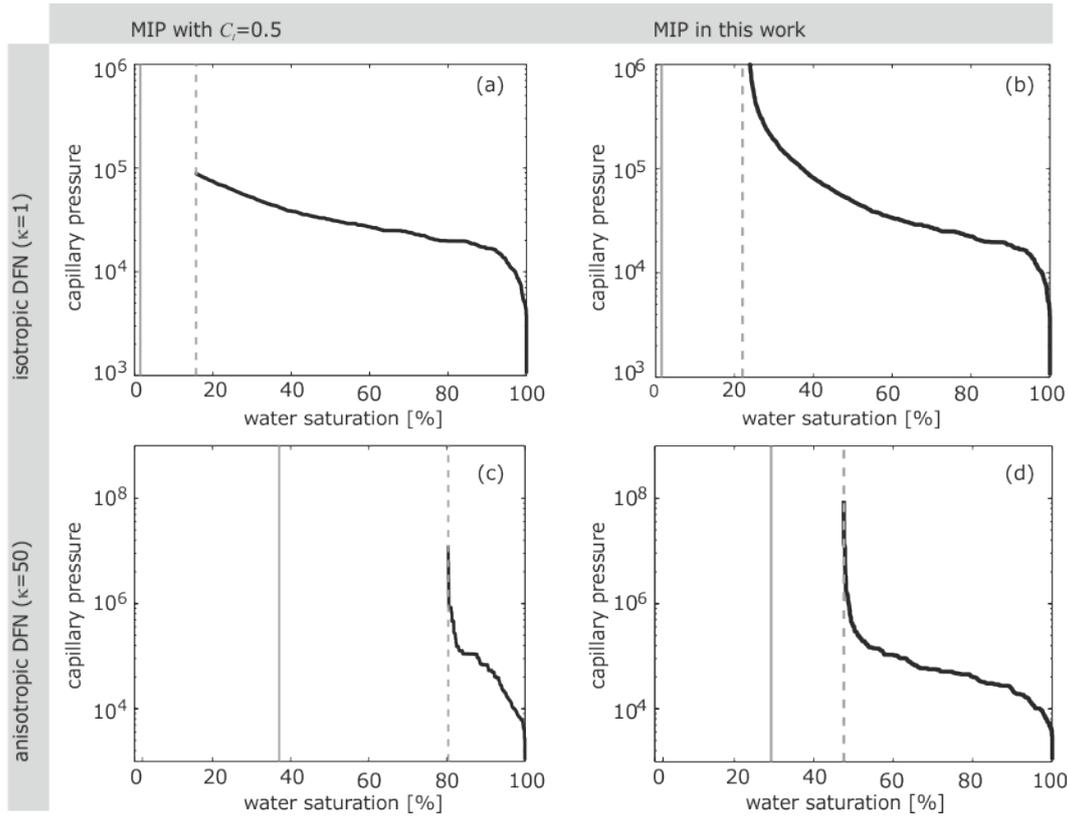

Fig. 11. Capillary pressure-water saturation relationship after full invasion for trapping MIP on the isotropic DFN ($\kappa$=1) and anisotropic DFN ($\kappa$=50). (a) and (c)Trapping MIP ($C_I$ = 0.5). (b) and (d) new developed MIP. The total amount of trapped water is marked by a dashed line. The amount of trapped water of pruned "dead end" fractures that are not invadable is indicated by the solid line.

## 5. Summary and conclusions

We introduced a multi-scale framework that maps flow in heterogeneous fractures from FE calculations on computationally efficient, hydraulically corresponding artificial channel networks (ACNs) that are used to calculate system specific scaling relations for fracture sizes and fracture interactions. The utilized algorithms for channel identification and characterization are based on robust image processing routines and network theory. Using these distributions for the channel properties ACNs are generated. It was shown, that coarse grained, vectorized ACNs capture the hydraulic properties of single fractures with reasonable accuracy, but with highly reduced information. To quantify the intra- and inter-fracture scaling behavior of transport properties of fractures, parameter variations on the ACN are performed. These are the path length, inclination as well as intersection length of the entrance and outlet of the fracture. Discrete fracture networks can now be improved by considering not just a uniform aperture for a fracture, but by assigning equivalent apertures between fracture intersections, as well as entry apertures that consider intersection lengths and spatial relations. It is shown for a previously studied example, how these relations change the invasion of immiscible fluids into fully saturated, large fracture networks by trapping MIP (Wettstein et al. 2012).

The entire procedure is based on the assumption that the origin of scale effects in fracture networks stems from the fracture flow that is mainly localized inside a channel network with characteristic, scale dependent network properties like a characteristic length scale among others. The study was limited to spatially uncorrelated fracture surfaces with power-law disorder. Also a degree of disorder was chosen that leads to clear flow localization in channels for Stokes flow. However different aperture fields, higher Reynolds numbers or parallel plate flow for non-channelized portions can be considered with small additional effort to combine the effects of heterogeneity in the single fracture scale on the network scale.

When additional knowledge about the finer scale is incorporated on larger scales, in general more physical results are obtained. We demonstrated how an efficient two-scale approach for invasion and flow in fracture networks can work with a model of only one input parameter that controls the heterogeneity of the aperture size distribution. We compare results from the two-scale MIP model with the previously developed MIP model for artificial fracture networks, since absolute data for verification is abundant. Since the presented model considers scale effects originating form disordered surface roughness such as intersections and flow path lengths of underlying channel networks, it can be seen as the improved and more accurate model to predict transport properties of the fractured rocks.

## Acknowledgment

Financial support from nagra, ETH Zurich, and the European Research Council (ERC) under Advanced Grant No. 319968-FlowCCS is gratefully acknowledged. The authors are thankful to Paul Marshall (nagra) and Bill Lanyon (Fracture Systems Ltd.), for fruitful discussions and ongoing support.

## Appendix A. Width and Equivalent Aperture Size of Channels

To measure the width of the channels, circles with different radii are drawn to find the intersection between channel wall and circles. Their centers are placed on the skeleton (see Fig. A1). The mean of the flow rate profile for different cross sections inside one typical channel is normal distributed, hence characterized by variance and amplitude. It is assumed that the flow rates in the boundaries of the channels are less that 5% of the maximum flow rate inside the channels. Circles with increasing radius are drawn and simultaneously the corresponding values of the flow rates for each radius are checked until the criterion is satisfied. The final diameter of the circle is taken as the width of the channel.

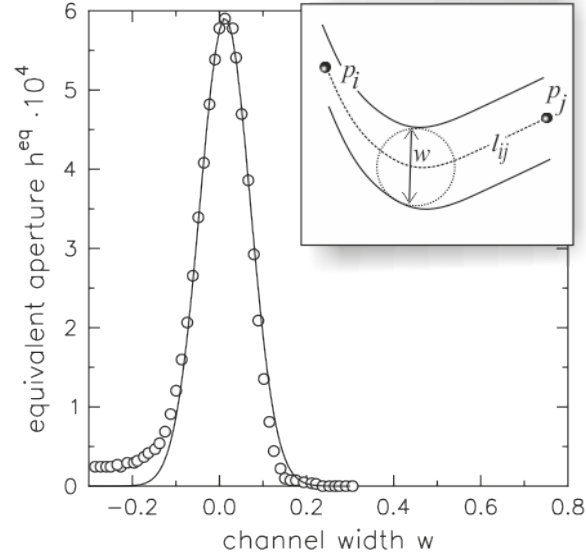

**Fig. A1. Flow rate profile of channels with fitted Gaussian curve. Representation of a channel (inset). The dashed line indicates the skeleton while the double sided arrow point at the width of the channel.**

Using the cubic law, the equivalent aperture size $h_{ij}^{eq}$ for a channel from node $i$ to node $j$ is computed by

$$h_{ij}^{eq} = \sqrt[3]{\frac{12 q_{ij} l_{ij}}{w(p_i - p_j)}}, \qquad (A1)$$

where $q_{ij}$ is the flow rate inside the channel between node $i$ and $j$; $p_i$ and $p_j$ are the pressure at node $i$ and $j$; $l$ and $w$ are length and width of the channel, respectively.

**Highlights:**

- A modified percolation model is developed by considering heterogeneity on the scale of the individual fractures to incorporate size effects.
- Individual fractures are coarse grained using an artificial channel network
- Equivalent and entry apertures depend on fracture intersection angle, path length and intersection length.
- Modifications are applied to simulate invasion percolation processes through two large isotropic and anisotropic discrete fracture networks.

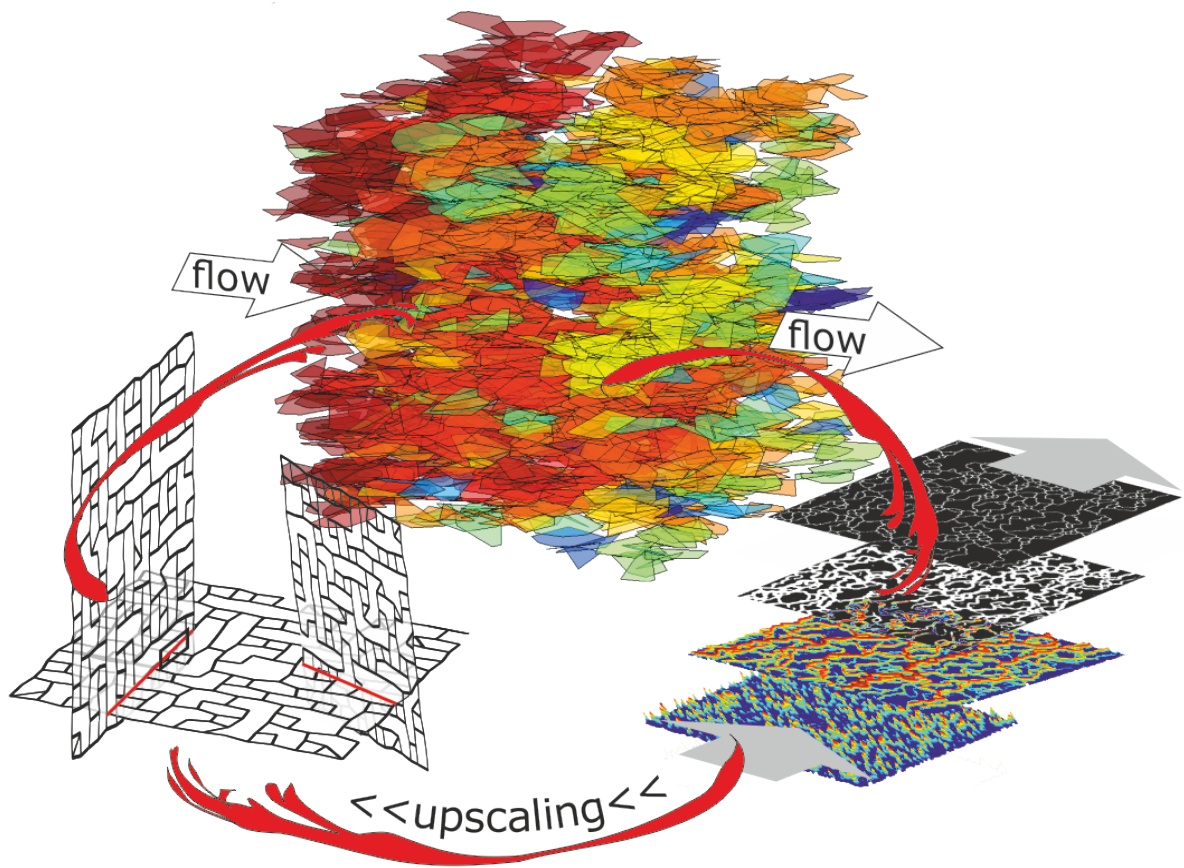